\documentstyle[12pt,epsf]{article}
\chardef\letterchar=11
\chardef\otherchar=12
\chardef\eolinechar=5
\ifx\selectfont\undefined%
  \def\mathrm#1{{\rm #1}}
\fi
\def\ifmath#1{\relax\ifmmode #1\else $#1$\fi}%
\let\ifmathx0
\def\fixmath{\def\ifmath{\noexpand\ifmathx}}%

\newcount\hrs\newcount\minu\newcount\temptime
\def\hm{\hrs=\time \divide\hrs by 60 \minu=\time\temptime=\hrs
\multiply\temptime by 60%
\advance\minu by -\temptime
\ifnum\minu<10 \let\zerofill=0\else \let\zerofill=\relax\fi
 \the\hrs:\zerofill\the\minu}%
\def\lapprox{\ifmath{\sim\kern-1em\raise 0.65ex\hbox{$<$}}}
\def\rapprox{\ifmath{\sim\kern-1em\raise 0.65ex\hbox{$>$}}}
\def\Zo{\ifmath{\mathrm {Z}}}

\def\Zo{\ifmath{\mathrm {Z}}}

\def\xb{\ifmath{\raise.4ex\hbox{$\chi$}_{_{\mathrm{B}}}}}
\def\x{\raise.4ex\hbox{$\chi$}}

\def\chino{\ifmath{\mathchoice%
     {\displaystyle\raise.4ex\hbox{$\displaystyle\tilde\chi$}}%
        {\textstyle\raise.4ex\hbox{$\textstyle\tilde\chi$}}%
      {\scriptstyle\raise.3ex\hbox{$\scriptstyle\tilde\chi$}}%
{\scriptscriptstyle\raise.3ex\hbox{$\scriptscriptstyle\tilde\chi$}}}}

\let\etaa=\eta
\def\eta{\ifmath{\etaa}}%
\let\iotaa=\iota%
\def\iota{\ifmath{\iotaa}}%
\def\TeV{\ifmmode \hbox{\rm\ Te\kern -0.1em V}\else
                  \hbox{\mathrm{Te\kern -0.1em V}}\fi}%
\def\GeV{\ifmmode \hbox{\rm\ Ge\kern -0.1em V}\else
                  \hbox{\mathrm{Ge\kern -0.1em V}}\fi}%
\def\MeV{\ifmmode \hbox{\rm\ Me\kern -0.1em V}\else
                  \hbox{\mathrm{Me\kern -0.1em V}}\fi}%
\def\keV{\ifmmode \hbox{\rm\ ke\kern -0.1em V}\else
                  \hbox{\mathrm{ke\kern -0.1em V}}\fi}%
\def\eV{\ifmmode \hbox{\rm\ e\kern -0.1em V}\else
                 \hbox{\mathrm{e\kern -0.1em V}}\fi}%
\def\mrad{\ifmmode \hbox{\rm\ mrad}\else
                 \hbox{\mathrm{mrad}}\fi}%

\def\mum{\ifmmode \hbox{\rm $\mu$\kern -0.1em m}\else
                  \hbox{\mathrm{$\mu$\kern -0.1em m}}\fi}%
\def\MM{\ifmmode \hbox{\rm m \kern -0.2em m}\else
                  \hbox{\mathrm{m\kern -0.2em m}}\fi}%
\newbox\boxsqbox
\newdimen\boxsize\boxsize=1.2ex%
\def\boxop{%
\setbox\boxsqbox=\vbox{\hrule depth0.8pt width0.8\boxsize height0pt%
                       \kern0.8\boxsize
                       \hrule height0.8pt width0.8\boxsize depth0pt}%
           \hbox{%
           \vrule height1.0\boxsize width0.8pt depth0pt%
           \copy\boxsqbox
           \vrule height1.0\boxsize width0.8pt depth0pt\kern1.5pt}}%
\def\pmb#1{\setbox0=\hbox{$#1$}
  \kern-.025em\copy0\kern-1.0\wd0%
  \kern.05em\copy0\kern-1.0\wd0%
  \kern-.025em\raise.0433em\box0}%
\newdimen\dkwidth
\def\dk{%
   \dkwidth=\baselineskip
   {\def\to{\rightarrow}
   \kern 3pt%
   \hbox{%
      \raise 3pt%
      \hbox{%
         \vrule height 0.8\dkwidth width 0.7pt depth0pt%
      }%
      \kern-0.4pt%
      \hbox to 1.5\dkwidth{%
         \rightarrowfill
      }%
   \kern0.6em%
   }}%
}%
\def\eqalign#1{%
   \,
   \vcenter{%
      \openup\jot\m@th
      \ialign{%
         \strut\hfil$\displaystyle{##}$&&$%
         \displaystyle{{}##}$\hfil\crcr#1\crcr%
      }%
   }%
   \,
}%
\newcommand {\zfi}{$_{Z\!F}\!I\!^{\textstyle T}\!\!T\!\!_{{\textstyle E}\!R}$}

\newcommand{\csm}{$\cal SM$}

\newcommand{\cmi}{$\cal MI$}
\def\ta1{\ifmath{\tau \rightarrow {\mathrm a_1} \nu}}

\newcommand{\bc}{\begin{center}}
\newcommand{\ec}{\end{center}}
\newcommand{\bq}{\begin{equation}}
\newcommand{\eq}{\end{equation}}
\newcommand{\ba}{\begin{eqnarray}}
\newcommand{\ea}{\end{eqnarray}}
\newcommand{\ban}{\begin{eqnarray*}}
\newcommand{\ean}{\end{eqnarray*}}

\newcommand{\nll}{\nonumber \\}

\begin{document}
\thispagestyle{empty}
\onecolumn
\begin{flushleft}
{\tt
DESY 94-253
\\
hep-ph/
\\
December 1994
}
\end{flushleft}
\vspace*{1.0cm}
\begin{center}
{\LARGE
The line shape of the $Z$ boson
\footnote{Talk held at the 28$^{th}$
Int. Symp. Ahrenshoop on the Theory of Elementary Particles,
Wendisch-Rietz/Brandenburg, 30 Aug -- 3 Sept 1994, to appear in the
proceedings.}
}
\vspace{1.0cm}
\\
\noindent
\Large
Tord~Riemann
\footnote{email: riemann@ifh.de}
\vspace{.4cm}
\\
\noindent
\normalsize
{\Large \em
DESY -- Institut f\"ur Hochenergiephysik
\vspace{1mm}
\\
Platanenallee 6, D-15738 Zeuthen, Germany
\vspace{9mm}
}
\thispagestyle{empty}

\noindent
{\large
{Abstract
\\}
}
\end{center}
\noindent
At LEP~1,
cross sections and cross section asymmetries may be analysed
model independently.
Cross sections depend on four, asymmetries on two free
parameters.
As an example, I discuss the model independent $Z$ boson mass
determination from the $Z$ line shape and compare it to the
Standard Model approach.
\normalsize
\vspace*{1.0cm}
\section {
Introduction
\label {sec1}
}
The precision measurements of the weak neutral current reaction
\ba
e^+ e^- \rightarrow (\gamma, Z) \rightarrow f^+ f^- (n \gamma)
\label{e0}
\ea
are being performed at LEP~1 and SLC since 1989.
Until summer 1994, the following event samples have been
collected~\cite{schaile}:
\begin{itemize}
\item
$e^+ e^- \rightarrow {\bar q} q$: 7.1 Mio. (LEP~1)
\item
$e^+ e^- \rightarrow {\bar l} l$: 0.8 Mio. (LEP~1)
\item
$e^+ e^- \rightarrow$ all\footnote{Excluding Bhabha events.}:
0.05 Mio. (SLC)
\end{itemize}
{}From these data, one may derive the weak neutral current parameters
with unprecendented precision.
An unbiased interpretation of data becomes a highly
nontrivial task since data have to be understood as a
result not only of $e^+e^-$ scattering with one particle exchange
but also of radiative corrections.
The calculation of the latter may be, to some approximation, subdivided
into two different, separated problems:
the well understood and calculable QED and QCD corrections (including
real bremsstrahlung of photons and gluons and higher order corrections)
at one hand and, at the other, the model dependent virtual weak
corrections.
As long as the latter are small and may be absorbed into a small number
of parameters it will be reasonable to hope to interprete
data correctly without fixing the model.
In fact, there are three different popular approaches to the
$Z$ line shape data:
\begin{itemize}
\item
Standard Model (${\cal SM}$):
$\alpha_{em}, \alpha_{strong}, G_{\mu}, M_Z, m_t, M_H$;
\item
\csm\ plus New Physics (${\cal NP}$): assume the \csm\ parameters
as being known and determine the additional ones;
\item
Model independent (\cmi): $M_Z, \Gamma_Z$ and few others; see below.
\end{itemize}
An updated discussion of the \csm\ approach is being prepared
in~\cite{yr94}.
In the following I will concentrate on the model independent approach
to the $Z$ line shape
and the problem of a unique choice of parameters.
An especially interesting question is that of the minimal number of
free parameters.

The QED and QCD corrections may be taken into account with the following
convolution formula:
\ba
\sigma(s)
&=&\int\frac{ds'}{s}\sigma_0(s')\, \rho\left(\frac{s'}{s}\right)
{}~+~\int\frac{ds'}{s}\sigma_0^{int}(s,s')
\, \rho^{int}
\left(\frac{s'}{s}\right),
\label{sigqed2}
\ea
where the radiator $\rho$ describes initial and final state radiation,
including leading higher order effects and soft photon exponentiation,
while the second radiator $\rho^{int}$ takes into account the
initial-final state interference effects (see e.g.~\cite{bardin4}), which
are comparatively small (a few per mille)
but maybe not negligible in future.
The bulk of the QED corrections is absorbed in $\rho$, which is described
in detail at many places, e.g. in~\cite{yr94,bardin4,yr89} and in
references therein.
Aiming at an accuracy of per mille around
the $Z$ peak, the radiator $\rho$
is different
for cross sections, which are symmetric (like $\sigma^T$)
or anti-symmetric (like $\sigma^{FB}$) in the
scattering angle (see e.g.~\cite{bardin4};
the same holds true for $\rho^{int}$).
The QCD corrections (if any) are traditionally included as factors
to the basic, elementary cross section $\sigma_0$; see e.g.~\cite{yr94}.

If one considers the contribution from the initial-final state
interferences to be negligible (or prefers to calculate them in the \csm),
then the only unknown is the basic cross section as a function
of the invariant $s$.
Thus, the line shape problem has been reduced to the search for an ansatz
for $\sigma_0(s)$.
Under certain, weak assumptions one may e.g. derive from the data the
following five parameters from which the cross sections may be constructed
{}~\cite{schaile}:
\begin{eqnarray}
M_Z &=&  91.188\, 8 \pm 0.004\, 4 \GeV,
\nll
\Gamma_Z &=& 2.497\, 4 \pm 0.003\, 8 \GeV,
\nll
\sigma_0^{had} &=& 41.49 \pm 0.12 \,{\mbox{nb}},
\nll
R_l = \frac{\sigma_0^{had}}{\sigma_0^{lept}} &=& 20.795 \pm 0.040,
\nll
A_{FB,0}^{lept} &=& 0.017\, 0 \pm 0.001\, 6.
\end{eqnarray}
These parameters are considered to be primary parameters in contrast
to derived ones, e.g. the effective
leptonic weak neutral current couplings or the effective weak mixing angle
{}~\cite{schaile}~(for details see, again,~\cite{yr94}):
\begin{eqnarray}
(g_v^l)^2 &=& 0.001\, 44 \pm 0.000\, 14,
\nll
(g_a^l)^2 &=& 0.251\, 18 \pm 0.000\, 56,
\nll
\sin^{2}\vartheta_{W}^{eff}
\equiv \frac{1}{4}\left( 1-\frac{g_v^l}{g_a^l}\right)
&=& 0.231\, 07 \pm 0.000\, 90.
\label{sw1}
\end{eqnarray}
Another interesting  derived quantity is the invisible width of the
$Z$ boson, which may be derived from $\Gamma_Z$ and the observed partial
$Z$ widths;
or, alternatively, the number of light neutrino species
{}~\cite{schaile}:
\bq
\nonumber
N_{\nu} = 2.988 \pm 0.023.
\eq
\section{Model independent determination of the $Z$ mass}
The $Z$ boson mass determination is part of a global fit to a large
variety of $Z$ line shape data. It is dominated by the total hadronic
production cross section due to the high statistics of that reaction.
A sufficiently accurate ansatz for this cross section
is~\cite{riemann,MI,riemann3,bardin2}:
\ba
\sigma_0(s) =
\frac{4}{3} \pi \alpha^2
\left[ \frac{r^{\gamma}}{s} +
\frac {s\cdot r + (s - M_Z^2)\cdot j}
{\left|s-M_Z^2 + i s \Gamma_z/M_Z\right|^2}
\right],
\label{sigqed}
\ea
where the photon exchange parameter
$r^{\gamma}$ is assumed to be known.
The numerical value of the $Z$ mass is closely related to the peak
position $\sqrt{s_{\max}}$
of the $Z$ line shape; thus an estimate of the peak position
models problems connected with the mass measurement.
As a simplified ansatz let us use the following approximation of the
Breit-Wigner function:
\ba 
\sigma_0^Z(s) \sim
\frac {M_Z^2\cdot  r } {\left|s-M_Z^2 +i M_Z \Gamma_Z\right|^2}.
\ea

The bulk of the corrections is due to initial state radiation and may be
well described by the following formula~\cite{nicrosini} (see
also~\cite{bardin4} and the discussion in~\cite{yr89}):
\ba
\rho(z) &=& \beta (1-z)^{\beta-1}\delta^{S+V}+\delta_1^H+\delta_B^H,
\\
\beta &=&
 \frac{2\alpha}{\pi} \left(L-1\right), 
{}~~~\mbox{with}~~~
L = \ln\frac{s}{m_e^2},
\\
\delta^{S+V} &=&
1
+\frac{\alpha}{\pi} \left[\frac{3}{2} L +2\zeta(2) -2 \right]
+~ \left( \frac{\alpha}{\pi} \right)^2
\Biggl\{
\left[\frac{9}{8}-2\zeta(2)\right]L^2
+\left[ -\frac{45}{16}+\frac{11}{2}\zeta(2)+3\zeta(3)\right]L
\nll &&+~
\left[ -\frac{6}{5}\zeta(2)^2-\frac{9}{2}\zeta(3)-6\zeta(2)\ln2
+\frac{3}{8}\zeta(2)+\frac{19}{4}\right]
\Biggr\},
\\
\delta_1^H &=& -\frac{\alpha}{\pi} (1+z) \left(L-1\right),
\\
\delta_B^H &=&
\left( \frac{\alpha}{\pi} \right)^2 \frac{1}{2} \left(l-1\right)^2
\left\{ (1+z)\left[3\ln z-4\ln(1-z)\right]-\frac{4}{1-z}\ln z -5-z
\right\},
\ea
with $z=s'/s$.
The corresponding
peak shift, which has to be taken into account in order to correctly
determine the $Z$ mass is known since
long~\cite{lep2}:
\ba
\sqrt{s_{\max}} - M_Z &=&
\delta_{QED} =
\frac{\pi}{8} \beta
\left(1+\delta^{S+V}\right) \Gamma_Z
+ \,\mbox{small corrections}
{}~\approx~ 90 \,\mbox{MeV},
\ea
where $\delta^{S+V}$ describes virtual and soft real photon emission.

Improving the Breit-Wigner function by a replacement of $M_Z^2\cdot r$ by
$s\cdot r$ and of the $Z$ width in the denominator by an energy dependent
width function as preferred by the LEP collaborations,
the peak shift gets modified as follows:
\ba
\sqrt{s_{\max}} - M_Z &=&
\delta_{QED} \oplus
\frac{1}{4} \frac{\Gamma_Z^2}{M_Z}
\ominus
\frac{1}{2}\frac{\Gamma_Z^2}{M_Z}.
\ea
If these shifts are neglected, they have to be condered as systematic
errors.
The importance of the second of the modifications has been stressed first
in~\cite{berends, riemann2}. It amounts to
$-\Gamma_Z^2/(2M_Z)=-34$ MeV.

The influence of the $\gamma Z$ interference on the $Z$ mass determination
has been observed some time ago; see e.g.~\cite{riemann}. It leads to an
additional modification of the the $Z$ peak position:
\ba
\sqrt{s_{\max}} - M_Z &=&
\delta_{QED} \oplus
\frac{1}{4} \frac{\Gamma_Z^2}{M_Z} \left(1+\frac{j}{r}\right)
\ominus
\frac{1}{2}\frac{\Gamma_Z^2}{M_Z}.
\ea
The interesting point is the following.
Neglecting this interference (setting $j$=0) leads to an erraneous
systematic shift of the $Z$ mass of 17~MeV$\otimes(j/r)$.
If one wants to take into account the $j$, a model for its
prediction is needed.

{\it
The $Z$ line shape has four free parameters per channel:
$M_Z, \Gamma_Z, r, j$.
 If $j$ is not fixed but treated as a parameter
of a \cmi\
line shape fit, the $Z$ mass gets an additional error of $\Delta M_Z =
\pm 8 $ MeV/expt.
This uncertainty could be removed by a dedicated running of LEP~1 at
energies away from the peak~\cite{kirsch,grunewald}.
}
\section{\cmi\ approach to asymmetries}
For the description of cross section asymmetries, an analogue
to~(\ref{sigqed2}) may be used~\cite{riemann3}:
\ba
\label{eqn:mi_asy}
{\cal A}(s) = \frac{\sigma_A(s)}{\sigma_{T}(s)},
\ea
where subscript $A$ stands for the type of asymmetry, e.g. $A=FB$ for the
forward backward asymmetry, and both cross sections
$\sigma_0^A$ and
$\sigma_0^T$
have the form~(\ref{sigqed}), with different parameters $r^A, j^A$.
Around the \Zo\ resonance all asymmetries take an extremely simple form:
\ba
{\bar{\cal A}}_A(s)
&=&
{\bar A}_0^A
+ {\bar A}_1^A \left(\frac{s}{M_Z^2} - 1 \right),
\label{e24}
\ea
where the coefficients at the right hand side depend on $r^A, j^A$ and
on the QED corrections.
In~\cite{riemann,riemann3} it is explained in detail that the coefficient
${\bar A}_0^A \approx r^A/r^T$ is for $A=FB$ nearly (and for $A=LR,pol$
strictly)
independent of QED corrections
while
${\bar A}_1^A \approx C_{QED}^A(s)(j^A/r^A-j^T/r^T)
{\bar A}_0^A$ reflects effects from the radiative tail and is responsible
for deviations of the asymmetry from being constant ($j\neq0$) and from
a straight line ($C_{QED}\neq \mbox{const.}$); see
figure~1~\cite{riemann3}.

\def\swid{0.5\textwidth}
\begin{figure}[thbp]
\begin{center}
\end{center}
\caption[foo]{
\it
The forward-backward asymmetry for the process $e^+ e^- \rightarrow \mu^+
\mu^-$ near the \Zo\ peak.
\label{asy}
}
\end{figure}

{\it
Any asymmetry at LEP~1 is fully described by two
free parameters.
The \cmi\ approach
allows to determine the contribution of the $\gamma Z$ interference
to the asymmetries.
It will get more importance when more precise data will be available
as is expected in the near future.
Maybe it will help to understand the origin of the
discrepancy between the LEP~1
measurements and the SLD determination of $A_{LR}$~\cite{mfero},
which corresponds to
$\sin^2 \vartheta_W^{eff} = 0.229\, 4\pm 0.001\, 0$.
}
\section{The \csm\ approach}
The cross section $\sigma_0(s)$ in~(\ref{sigqed2})
may be parametrized in the \csm\ by four electroweak form
factors (\cite{bardin2} and references therein):
$\rho_{ef}, \kappa_e, \kappa_f,\kappa_{ef}$.
To a good approximation, these form factors are independent of
$s$ and $\cos\vartheta$ on which they depend in fact, and are even
universal (with small deviations) in the sense that they are flavor
independent (with the notable exclusion of $b$ quark production in view
of the large $t$ quark mass).
The effective couplings are related to them as follows:
$|a_f^{eff}|=\sqrt{\rho_{ef}}$,
$v_f^{eff}=a_f^{eff}(1-4|Q_f|s_W^{2,eff})$,
$s_W^{2,eff}=\kappa_f \sin^2
\vartheta_W$, $v_{ef}\approx v_e^{eff} v_f^{eff}$.
For a detailed discussion I refer to~\cite{yr94}.
The $Z$ exchange contribution to the cross section and to two of the
asymmetries is at the $Z$ peak:
\ba
\sigma_0^Z(M_Z^2) &\sim&
\left(1+ |v_e|^2+|v_f|^2+|v_{ef}|^2\right),
\\
{\cal A}_{FB}^Z(M_Z^2) &\approx& \frac{3}{4}
\frac{ \Re e \left[ v_e v_f^* + v_{ef}\right] }
     {1+ |v_e|^2+|v_f|^2+|v_{ef}|^2} \sim \frac{3}{4}
{\cal A}_e{\cal A}_f,
\\
{\cal A}_{LR}^Z(M_Z^2) &=&
\frac{ \Re e \left[ v_e  + v_{ef} v_f^* \right] }
     {1+ |v_e|^2+|v_f|^2+|v_{ef}|^2} \sim
{\cal A}_e,
\ea
with ${\cal A}_f=2v_fa_f/(v_f^2+a_f^2)$ and $|a_f^0|=1$.
I is easy to find relations between
the parameters of the \cmi\ approach and the parameters of the \csm.

As mentioned in the Introduction, one of the unknown parameters in the
\csm\ is the $t$ quark mass.
A global fit to the $Z$ line shape data yields~\cite{grunewald}:
\ba
M_Z &=&  91.188\, 7 \pm 0.004\, 4 \GeV,
\nll
m_t &=& 173 \pm 13 \pm 20~\mbox{(H)}~\mbox{GeV},
\nll
\alpha_{strong}(M_Z^2) &=& 0.126 \pm 0.005 \pm 0.002~\mbox{(H)},
\ea
where `H' indicates the estimated Higgs boson uncertainty.
The corresponding effective weak mixing angle is~\cite{schaile}:
\ba
\sin^2 \vartheta_W^{eff} = 0.232\, 2 \pm 0.000\, 4
\pm 0.000\, 2~\mbox{(H)}.
\ea
The $t$ quark mass
value has to be compared to the result of the direct $t$ quark search
at Fermilab with evidence for $m_t = 174 \pm 10 \pm 13$
GeV~\cite{fermilabtop}.

{\it
All the \csm\ determinations agree well with the findings of the \cmi\
approach, which are quoted above, and with the evidence of CDF for the
$t$ quark.
Besides QED and QCD higher order corrections,
the fermionic and bosonic
weak one loop corrections have to be taken into account.
Although leading higher order weak corrections play some role it is not
clear
whether a complete two loop calculation will be needed finally.
The Higgs boson mass cannot be estimated so far.
\\

{}From the agreement of the \cmi\ and the \csm\ fits one has to conclude
that there are no indications for New Physics presently.
\\}

{\bf Acknowledgement}
\\
I would like to thank D.~Bardin for a careful reading of the manuscript.

\end{document}